\documentclass{article}%
\usepackage{amsmath}
\usepackage{amsfonts}
\usepackage{amssymb}
\usepackage{graphicx}%
\begin{document}

\title{On unitary evolution and collapse \\in Quantum Mechanics}
\author{Francesco Giacosa\\\textit{Institute for Theoretical Physics, }\\\textit{J. W. Goethe University, Max-von-Laue-Str. 1, 60438, }\\\textit{Frankfurt am Main, Germany}}
\maketitle

\begin{abstract}
In the framework of an interference setup in which only two outcomes are
possible (such as in the case of a Mach-Zehnder interferometer), we discuss in
a simple and pedagogical way the difference between a standard, unitary
quantum mechanical evolution and the existence of a real collapse of the wave
function. This is a central and not-yet resolved question of Quantum Mechanics
and indeed of Quantum\ Field Theory as well. Moreover, we also present the
Elitzur-Vaidman bomb, the delayed choice experiment, and the effect of
decoherence. In the end, we propose two simple experiments to visualize
decoherence and to test the role of an entangled particle.

\end{abstract}

\section{Introduction}

Quantum Mechanics (QM) is a well-established theoretical construct, which
passed countless and ingenious experimental tests \cite{sakurai}. Still, it is
renowned that QM has some puzzling features
\cite{omnes,100,bell,bassighirardi,penrosebook}: are macroscopic
distinguishable superpositions (Schr\"{o}dinger-cat states) possible or there
is a limit of validity of QM? Do measurements imply a non-unitary
(collapse-like) time evolution or are they also part of a unitary evolution?
In the latter case, should we simply accept that the wave function splits in
many branches (i.e., parallel worlds), which decohere very fast and are thus
independent from each other? It is important to stress that these issues are
not only central in nonrelativistic QM but apply also in relativistic Quantum
Field Theory. Namely, the generalization to quantized fields does not modify
the role of measurements.

In this work we discuss in a introductory way some of the questions mentioned
above. We study the quantum interference in an idealized two-slit experiment
and we analyze the effect that a detector measuring \textquotedblleft which
path has been taken\textquotedblright\ has on the system. In particular, we
shall concentrate on the collapse of the wave function, such as the one
advocated by collapse models
\cite{bassighirardi,penrosebook,bassi,grw,pearle,penrose,diosi,singh} and show
which are the implications of it.

Variants of our setup also lead us to the presentation of the famous
Elitzur-Vaidman bomb \cite{bomb} and to delayed choice experiments
\cite{dc1,dc2}. Thus, we can describe in a unified framework and with simple
mathematical steps (typical of a QM course) concepts related to modern issues
and experiments of QM.

Besides the pedagogical purposes of this work, we also aim to propose two
experiments (i) to see decoherence at work in an interference setup with only
two possible outcomes and (ii) to test the dependence of the interference on
an idler entangled particle.

\section{Collapse vs no-collapse: no difference?}

\subsection{Interference setup}

We consider an interference setup as the one depicted in Figs. 1 and 2. A
particle $P$ flies toward a barrier which contains two `slits' and then flies
further to a screen $S$. Usually in such a situation there is a superposition
of waves which generates on the screen $S$ many maxima and minima. We would
like to avoid this unnecessary complication here but still use the language of
a double-slit experiment in which a sum over paths is present. To this end, we
assume that the particle can hit the screen in two points only, denoted as $A$
and $B$, see the discussion below. All the issues of QM can be studied in this
simplified framework. We assume also to `sit on' the screen $S$: when the
particle hits $A$ or $B$ we `see' it.

%

\begin{figure}
[ptb]
\begin{center}
\includegraphics[
height=2.4396in,
width=4.7712in
]%
{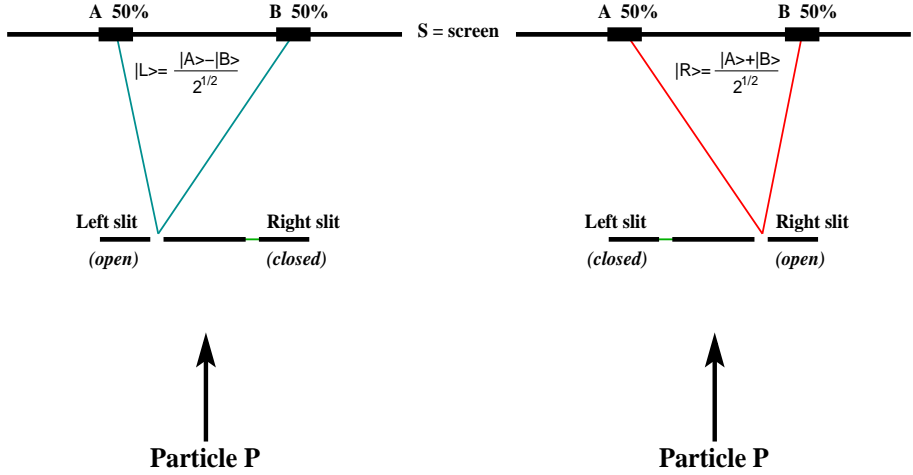}%
\caption{Hypothetical experiment with only two possible outcomes ($A$ and
$B$). Left: only the left slit is open. Right: only the right slit is open.
Note, each slit is not a simple hole but acts as a filter which projects the
particle either to a trajectory with endpoint $A$ or to a trajectory with
endpoint $B$.}%
\end{center}
\end{figure}

First, we consider the case in which only the left slit is open (Fig.~1, left
side). In order to achieve our goal, the slit is actually not a simple hole in
the barrier (out of which a spherical wave would emerge) but a more
complicated filter which projects the particle either to a straight trajectory
ending in $A$ or to a straight trajectory ending in $B,$ see Fig.~1. In the
language of QM, this situation amounts to a wave function $\left\vert
L\right\rangle $ associated to the particle which has gone through the left
slit, which is \emph{assumed} to be:%
\begin{equation}
\left\vert L\right\rangle =\frac{1}{\sqrt{2}}\left(  \left\vert A\right\rangle
-\left\vert B\right\rangle \right)  \text{ .}\label{l}%
\end{equation}
Then, by simply using the Born rule (i.e., by squaring the coefficient
multiplying $\left\vert A\right\rangle $ or $\left\vert B\right\rangle $), we
predict that the particle ends up either in the endpoint $A$ with probability
$50\%$ or in the endpoint $B$ with probability $50\%$. This is indeed what we
measure by repeating the experiment many times. As we see, the probability is
-for us observer on the screen $S$- a fundamental ingredient of QM, which
however enters only in the very last step, i.e. when the measurement comes
into the game. The state $\left\vert L\right\rangle $ is an equal
(antisymmetric) superposition of $\left\vert A\right\rangle $ and $\left\vert
B\right\rangle $, but in a single experiment we do \emph{not} find a pale spot
on $A$ and a pale spot on $B$: we always find the particle either fully in $A$
or in $B.$ It is only after many repetitions of the experiment that we realize
that the outcome $A$ and the outcome $B$ are equally probable.

If only the right slit is open (Fig.~1, right side), we have a similar
situation in which only two trajectories ending in $A$ and in $B$ are present.
The wave function of the particle after having gone through the right slit is
denoted by $\left\vert R\right\rangle $ and is described by the orthogonal
combination to $\left\vert L\right\rangle $:$\ $
\begin{equation}
\text{ }\left\vert R\right\rangle =\frac{1}{\sqrt{2}}\left(  \left\vert
A\right\rangle +\left\vert B\right\rangle \right)  \text{ .}%
\end{equation}
Then, also in this case one finds the particle in $50\%$ of cases in $A$ and
$50\%$ in $B.$%

\begin{figure}
[ptb]
\begin{center}
\includegraphics[
height=3.6616in,
width=3.5267in
]%
{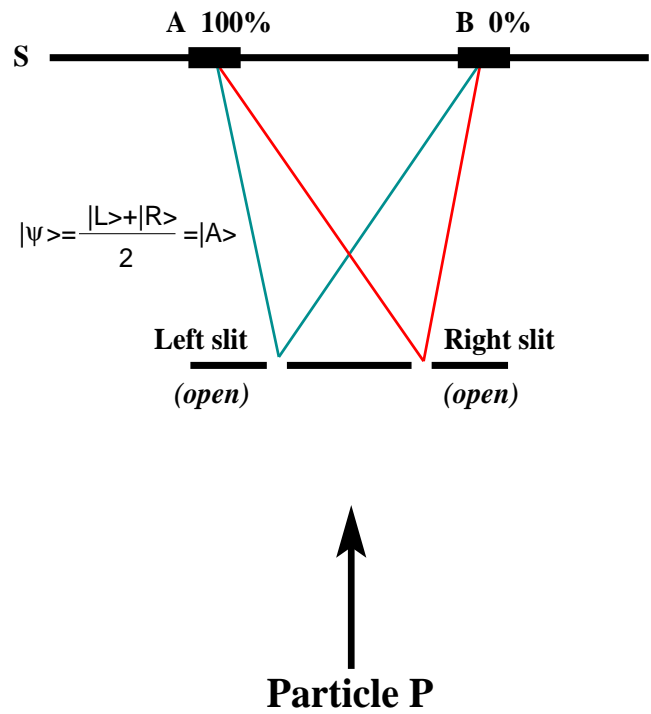}%
\caption{Same setup of Fig.~1, but now both slits are open: interference takes
place and all particles hit the screen in $A$.}%
\end{center}
\end{figure}

We now turn to the case in which both slits are open, see Fig.~2. The wave
function of the particle is assumed to be the sum of the contributions of the
two slits:%
\begin{equation}
\left\vert \Psi\right\rangle =\frac{1}{\sqrt{2}}\left(  \left\vert
L\right\rangle +\left\vert R\right\rangle \right)  \text{ ,} \label{s}%
\end{equation}
i.e. the contributions of both slits add coherently. A simple calculation
shows that
\begin{equation}
\left\vert \Psi\right\rangle =\left\vert A\right\rangle \text{ ,}%
\end{equation}
which means that the particle $P$ \emph{always} hits the screen in $A$ and
\emph{never} in $B.$ Namely, in $A$ we have a \emph{constructive}
interference, while in $B$ we have a \emph{destructive} interference. (Notice
that the points $A$ and $B$ are not equidistant from the two slits. However,
we take the two slits as being close to each other and the points $A$ and $B$
as being far from each other: the difference between the segments $LA$ and
$RA$ (and so between $LB$ and $RB$) is assumed to be negligible such that the
two contributions of the wave packet of the particle $P$ from the left and
right slit arrive almost simultaneously and the depicted interference effect
takes place).

In conclusion, we have chosen the language of a two-slit experiment because it
is the most intuitive. The price to pay is a slit acting as a filter and not
as a simple hole. However, one can easily build analogous setups as the one
here described by using photon polarizations, electron spins or equivalent
quantum objects, or by using a Mach-Zehnder interferometer, see details in
Sec. 2.3.3.

\subsection{Detector measuring the path}

As a next step we put a detector $D$ right after the two slits (both open).
$D$ measures through which hole the particle has passed, without destroying it
(see Fig.~3). We analyze the situation in two ways: first, by assuming the
collapse of the wave function as induced by $D$ and, second, by studying the
entanglement of the particle with the detector. Note, we still assume that we
sit on (or watch) the screen $S$ only, but we are not directly connected to
the detector $D.$

\bigskip

\emph{Collapse}: In this case we assume that the detector $D$ generates a
collapse of the wave function. Suddenly after the interaction with $D,$ the
state of the particle $P$ collapses into $\left\vert L\right\rangle $ with a
probability of $50\%$ or into $\left\vert R\right\rangle $ with a probability
of $50\%$. Then, the state is described by either $\left\vert L\right\rangle $
or $\left\vert R\right\rangle $, but not any longer by the superposition of
them. As a consequence, we have in half of the cases a situation analogous to
having only the left slit open and in the other half to having only the right
slit open.

What we will then see on the screen $S$? The probability to find the particle
in $A$ is given by%
\begin{equation}
P[A]=P[L,A]+P[R,A]=\frac{1}{2}\cdot\frac{1}{2}+\frac{1}{2}\cdot\frac{1}%
{2}=\frac{1}{2}%
\end{equation}
where $P[L,A]=1/4$ is the probability that the detector $D$ has measured the
particle going through the left slit and then the particle has hit the screen
in $A.$ Similarly, $P[R,A]=1/4$ is the probability that the detector $D$ has
measured the particle going through the right slit before the latter hits $A.$
A similar description holds for $P[B]=1/2$ with%
\begin{equation}
P[B]=P[L,B]+P[R,B]=\frac{1}{2}\cdot\frac{1}{2}+\frac{1}{2}\cdot\frac{1}%
{2}=\frac{1}{2}\text{ .}%
\end{equation}

The collapse is obviously part of the standard interpretation of QM, in which
a detector is treated as a classical object which induces the collapse of the
quantum state. As a result, there is no interference on the screen $S$. As
renowned, the standard interpretation does not put any border between what is
a classical system and what is a quantum system. Nevertheless, one can
interpret the collapse postulate as an effective description of a physical
process. Namely, in theories with the collapse of the wave function, the
collapse is a real physical phenomenon which takes place when one has a
macroscopic displacement of the position wave function of the detector (or,
more generally, of the environment). In this framework, somewhere in between
the quantum world and the classical macroscopic world, a \emph{new} physical
process takes place which realizes the collapse: this could be, for instance,
the stochastic hit in the Ghirardi-Rimini-Weber model
\cite{bassighirardi,bassi,grw} or the instability due to gravitation in the
Penrose-Diosi approach \cite{penrosebook,penrose,diosi}. Neglecting details,
the main point is that such collapse theories realize physically the collapse
which is postulated in the standard interpretation and liberates it from
inconsistencies. Still, it is an open and well posed physical question if (at
least one of) such collapse theories are (is) correct.%

\begin{figure}
[ptb]
\begin{center}
\includegraphics[
height=2.7432in,
width=4.1649in
]%
{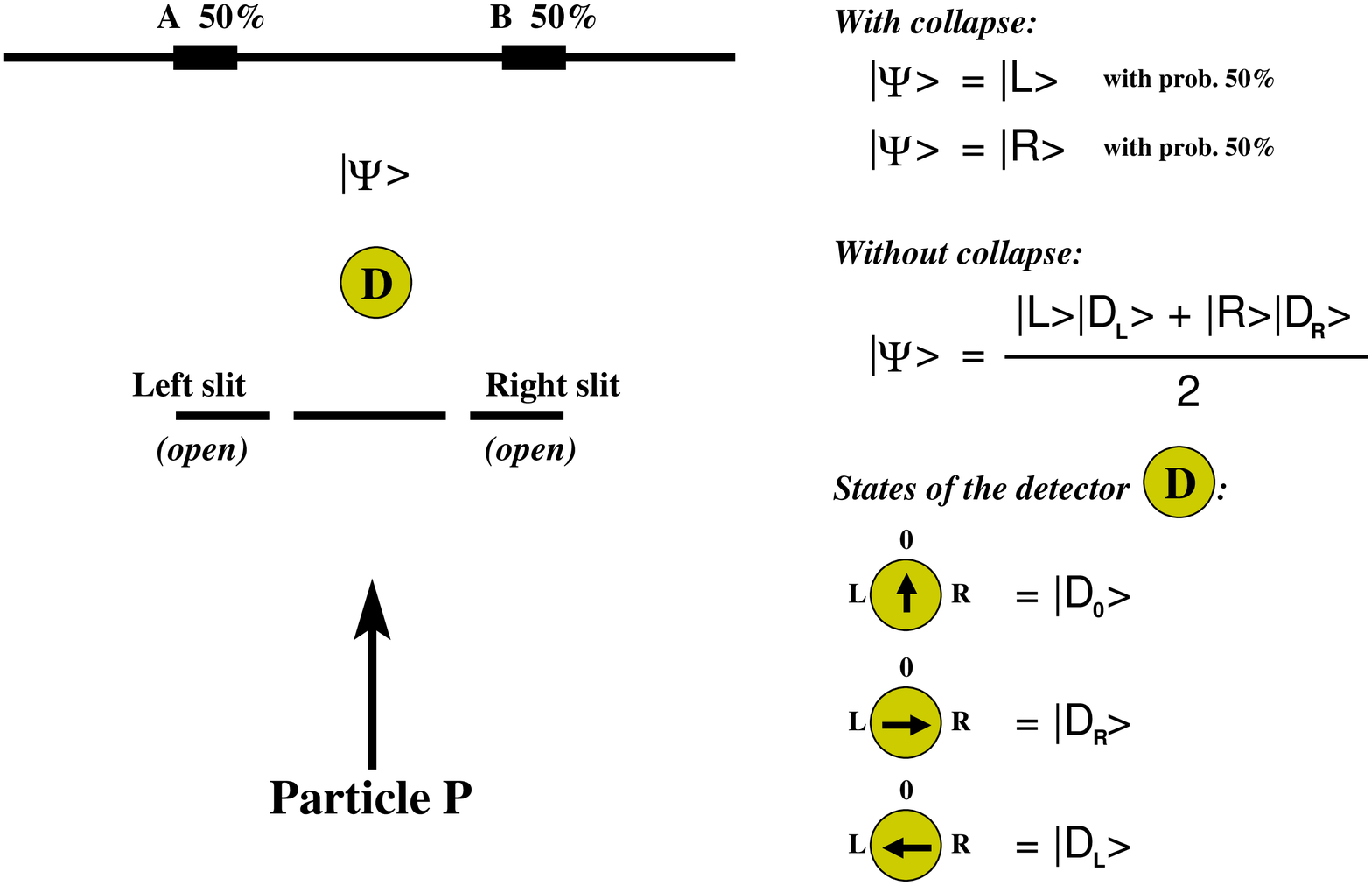}%
\caption{A detector which measures which slit the particle has gone through is
placed just after the slits. The wave functions for the collapse and
no-collapse scenarios are depicted.}%
\end{center}
\end{figure}

\bigskip

\emph{No-collapse: }In this case we do not assume that the detector $D$
generates a collapse of the wave function, but we enlarge the whole wave
function of the system by including also the wave function of the detector. We
assume that, prior to measurement, the detector is in the state $\left\vert
D_{0}\right\rangle $ (we can, for definiteness, think of a old-fashion
indicator which points to $0$, see Fig.~3). Then, when both slits are open,
the state of the whole system just after having passed through them but not
yet in contact with the detector $D$, is given by
\begin{equation}
\left\vert \Psi\right\rangle =\frac{1}{\sqrt{2}}\left(  \left\vert
L\right\rangle +\left\vert R\right\rangle \right)  \left\vert D_{0}%
\right\rangle \text{ .}%
\end{equation}
Then, the particle-detector interaction induces a (we assume very fast) time
evolution which generates the following state:%
\begin{equation}
\left\vert \Psi\right\rangle =\frac{1}{\sqrt{2}}\left(  \left\vert
L\right\rangle \left\vert D_{L}\right\rangle +\left\vert R\right\rangle
\left\vert D_{R}\right\rangle \right)  \text{ ,}%
\end{equation}
where $\left\vert D_{L}\right\rangle $ ($\left\vert D_{R}\right\rangle $)
describes the pointer of the detector pointing to the left (right). Thus, no
collapse is here taken into account, because the whole wave function still
includes a superposition of $\left\vert L\right\rangle $ and $\left\vert
R\right\rangle $, which, however, are now entangled with the detector states
$\left\vert D_{L}\right\rangle $ and $\left\vert D_{R}\right\rangle $, respectively.

An important point is that the overlap of $\left\vert D_{L}\right\rangle $ and
$\left\vert D_{R}\right\rangle $ is small:%
\begin{equation}
\left\langle D_{L}|D_{R}\right\rangle \simeq0\text{ ,}%
\end{equation}
to a very good degree of accuracy. To show it, let us ignore the rest of the
detector and the environment and concentrate on the pointer only, which is
assumed to be made of $N$ atoms, where $N$ is of the order of the Avogadro
constant. The atom $\alpha$ of the pointer is in a superposition of the type
$\left(  \psi_{L}^{\alpha}(\vec{x})+\psi_{R}^{\alpha}(\vec{x})\right)
/\sqrt{2}$, where $\psi_{L}^{\alpha}(\vec{x})$ ($\psi_{R}^{\alpha}(\vec{x})$)
is the wave function of the atom when the pointer points to the left (right).
We have:
\begin{equation}
\left\langle D_{L}|D_{R}\right\rangle =%
{\displaystyle\prod\limits_{\alpha=1}^{N}}
\int d^{3}x\left(  \psi_{L}^{\alpha}(\vec{x})\right)  ^{\ast}\psi_{R}^{\alpha
}(\vec{x})\text{ .}%
\end{equation}
The quantity $\int d^{3}x\left(  \psi_{L}^{\alpha}(\vec{x})\right)  ^{\ast
}\psi_{R}^{\alpha}(\vec{x})=\lambda_{\alpha}$ is such that $\left\vert
\lambda_{\alpha}\right\vert <1$. For a large displacement, $\lambda_{\alpha}$
is itself a very small number (small overlap), but the crucial point is to
observe that $\left\langle D_{L}|D_{R}\right\rangle $ is the product of many
numbers with modulus smaller then 1. Assuming that $\lambda_{\alpha}=\lambda$
for each $\alpha$ (each atom gets a similar displacement: this assumption is
crude but surely sufficient for an estimate), we get
\begin{equation}
\left\langle D_{L}|D_{R}\right\rangle \simeq\lambda^{N}\text{ ,}%
\end{equation}
which is extremely small for large $N.$ Even if we take $\lambda=0.99$ (which
is indeed quite large and actually overestimates the overlap of the wave
functions of an atom belonging to macroscopic distinguishable configuration),
we obtain
\begin{equation}
\left\langle D_{L}|D_{R}\right\rangle \simeq0.99^{N_{A}}\sim10^{-10^{21}}
\label{overlap}%
\end{equation}
which is tremendously small.

After having clarified the \emph{de facto} orthogonality of $\left\vert
D_{L}\right\rangle $ and $\left\vert D_{R}\right\rangle $, we rewrite the full
wave function of the system $\left\vert S\right\rangle $ as
\begin{equation}
\left\vert \Psi\right\rangle =\frac{1}{2}\left[  \left\vert A\right\rangle
(\left\vert D_{R}\right\rangle +\left\vert D_{L}\right\rangle )+\left\vert
B\right\rangle (\left\vert D_{R}\right\rangle -\left\vert D_{L}\right\rangle
)\right]  \text{ .}%
\end{equation}
Then, the probability to find the particle $P$ in $A$ is obtained (now by
using the Born rule, because we are observing the screen $S$):%
\begin{equation}
P[A]=P[L,A]+P[R,A]=\frac{1}{2}\cdot\frac{1}{2}+\frac{1}{2}\cdot\frac{1}%
{2}=\frac{1}{2}%
\end{equation}
where $P[L,A]=1/4$ is the probability that the system is described by
$\left\vert A\right\rangle \left\vert D_{L}\right\rangle $ and $P[R,A]=1/4$
the probability that it is described by $\left\vert A\right\rangle \left\vert
D_{R}\right\rangle .$ A similar situation holds for $P[B]=1/2.$ Thus, also in
this case the presence of $D$ causes the disappearance of interference.

The same result is obtained if we use the formalism of the statistical
operator, which is defined by $\hat{\rho}=\left\vert \Psi\right\rangle
\left\langle \Psi\right\vert $ (see, for instance, Refs.
\cite{sakurai,bassighirardi}). Upon tracing over the detector states
(environment states) the reduced statistical operator reads (we use here
$\left\langle D_{L}|D_{R}\right\rangle =0$):%

\begin{align}
\hat{\rho}_{red}  &  =\left\langle D_{L}\left\vert \hat{\rho}\right\vert
D_{L}\right\rangle +\left\langle D_{R}\left\vert \hat{\rho}\right\vert
D_{R}\right\rangle \nonumber\\
&  =\left(
\begin{array}
[c]{cc}%
\left\vert A\right\rangle  & \left\vert B\right\rangle
\end{array}
\right)  \left(
\begin{array}
[c]{cc}%
\frac{1}{2} & 0\\
0 & \frac{1}{2}%
\end{array}
\right)  \left(
\begin{array}
[c]{c}%
\left\langle A\right\vert \\
\left\langle B\right\vert
\end{array}
\right)  \text{ ,}%
\end{align}
where the diagonal elements represent $p[A]=p[B]=1/2$ respectively, while the
off-diagonal elements vanish in virtue of the (for all practical purposes)
orthogonality of $\left\vert D_{L}\right\rangle $ and $\left\vert
D_{R}\right\rangle $.

\bigskip

\emph{Sum up}$\emph{:}$ We find that, for us sitting on the screen $S$, the
\emph{very same outcome}, i.e. the absence of interference, is obtained by
applying the collapse postulate as an intermediate step due to the detector
$D$ or by considering the whole quantum state -including the detector $D$- and
by applying the Born rule \emph{only} in the very end. This equivalence holds
as long as the (anyhow very small) overlap of the detector states of
Eq.~(\ref{overlap}) is neglected (see also the related discussion in\ Sec. 3).
The question is then: do we need the collapse? The second calculation
(no-collapse) seems to answer us: `no, we don't'. In this respect, one has a
superposition of macroscopic distinct states, which coexist and are nothing
else but the branches of the Everett's or many worlds interpretation (MWI) of
QM \cite{everett}. Thus, assuming that no collapse takes place brings us quite
naturally to the MWI \cite{100,wheeler,dewitt,tegmark,notemwi}.

However, care is needed: in fact, the `no collapse' assumption is a general
statement and means also that there is no collapse when the particle $P$ hits
the screen $S$ (where \emph{our own} wave function is part of the game). Let
us clarify better this point by going back to the very first case we have
studied, in which only the left slit was open and no detector $D$ was present
(Fig.~1, left part). The wave function of the particle just before hitting the
screen is given by $\left\vert L\right\rangle =\left(  \left\vert
A\right\rangle -\left\vert B\right\rangle \right)  /\sqrt{2}$. But then, after
the hit and assuming no collapse, the whole wave function -including us, who
are the observers - reads:
\begin{align}
\left\vert \Psi\right\rangle  &  =\frac{1}{\sqrt{2}}\left\vert A\right\rangle
\left\vert \text{Screen recording }A\text{ and we observing }A\right\rangle
\nonumber\\
&  -\frac{1}{\sqrt{2}}\left\vert B\right\rangle \left\vert \text{Screen
recording }B\text{ and we observing }B\right\rangle \text{ .} \label{send}%
\end{align}
The question is why the coefficient in front of the vector
\[
\left\vert A\right\rangle \left\vert \text{Screen recording }A\text{ and we
observing }A\right\rangle
\]
tells us which is the \emph{subjective} probability of observing $A$ for the
observer (us) sitting on the screen. In other words, how does the MWI explain
the probabilities according to the Born rule? The Born rule seems to be an
additional postulate, which has to be put \textit{ad hoc} into it. This
situation is however not satisfactory, because the main idea of the MWI is to
eliminate the collapse from the description of the QM and consequently to
\emph{derive} the standard Born probabilities. Although there are attempts to
show that there is no need of postulating the Born rule in this context
\cite{dsw} (see also Ref. \cite{zurek}), no agreement has been reached up to
now \cite{bassighirardi,rae,hsu,commentmwi}. This is indeed an argumentation
in favour of the possibility that a collapse really takes place. Surely, `real
collapse' scenarios deserve to be studied theoretically and experimentally
\cite{bassighirardi,penrosebook,bassi}.

Note, up to now we did not mention the decoherence, see e.g.~Refs.
\cite{omnes,introdec,zurekrev,schlosshauer,schlosshauershort} and refs.
therein. This is possible because we have put a detector that makes a
measurement by evolving from the state $\left\vert D_{0}\right\rangle $ into
two (almost) orthogonal states $\left\vert D_{L}\right\rangle $ and
$\left\vert D_{R}\right\rangle $, but actually one can interpret this fast
change of the detector state as the result of a decoherence phenomenon. This
is however a rather peculiar decoherence, because we have prepared the
detector in a particular (low entropic) $\left\vert D_{0}\right\rangle $
state, which is `ready to' evolve into $\left\vert D_{L}\right\rangle $ and
$\left\vert D_{R}\right\rangle $ as soon as it interacts with the particle
$P$. In Sec. 3 we will describe what changes when the environment, instead of
the detector, is taken into account.

\subsection{Variants of the setup}

\subsubsection{The bomb}

A simple change of the setup allows us to present the famous Elitzur-Vaidman
bomb, first described in\ Ref. \cite{bomb} and then experimentally verified
in\ Ref. \cite{zeilinger}. We substitute the detector with a `bomb', which can
be activated by the particle $P$. We place the bomb only in front of the left
slit, see Fig.~4. This means that, if only the left slit is open, the bomb
explodes soon after the particle has gone through the slit. If, instead, only
the right slit is open, it doesn't explode. For definiteness and simplicity we
assume that the particle is not destroyed nor absorbed by the bomb.

Just as previously, we can interpret the experiment applying either the
collapse or by studying the whole wave function. In the collapse approach, the
bomb simply makes a measurement. When both slits are open the wave function,
before the interaction with the bomb, is given by $\left\vert \Psi
\right\rangle =\left(  \left\vert L\right\rangle +\left\vert R\right\rangle
\right)  /\sqrt{2}$: we will have an explosion in $50\%$ of cases and no
explosion in the remaining $50\%.$ Notice that in the second case the bomb is
doing a \emph{null} measurement. The very fact that the bomb does \emph{not}
explode means that the particle went to the right slit (we assume 100\%
efficiency in our ideal experiment). When the bomb explodes there is a
collapse into $\left\vert \Psi\right\rangle =\left\vert L\right\rangle $, when
it doesn't into $\left\vert \Psi\right\rangle =\left\vert R\right\rangle .$
Then, we have a situation which is very similar to the case of the detector
$D$ which we have studied previously: no interference on the screen $S$ is
observed, but we observe the particle in the endpoint $A$ and $B$ with
probability $1/2$ each.

If we do not assume the collapse of the wave function, the whole wave function
is given by (after interaction with the bomb)
\begin{align}
\left\vert \Psi\right\rangle  &  =\frac{1}{\sqrt{2}}\left(  \left\vert
L\right\rangle \left\vert B_{E}\right\rangle +\left\vert R\right\rangle
\left\vert B_{0}\right\rangle \right) \nonumber\\
&  =\frac{1}{2}\left[  \left\vert A\right\rangle (\left\vert B_{0}%
\right\rangle +\left\vert B_{E}\right\rangle )+\left\vert B\right\rangle
(\left\vert B_{0}\right\rangle -\left\vert B_{E}\right\rangle )\right]
\label{bomb}%
\end{align}
where $\left\vert B_{0}\right\rangle $ is the state describing the unexploded
bomb and $\left\vert B_{E}\right\rangle $ the exploded one. Obviously, as in
Eq.~(\ref{overlap}), we have $\left\langle B_{E}|B_{0}\right\rangle \simeq0.$
Again and just as before no interference is seen on $S$ but the two outcomes
$A$ and $B$ are equiprobable. Clearly, no difference between assuming the
collapse or not is found, but the interesting fact is that the non-explosion
of the bomb is enough to destroy interference.%

\begin{figure}
[ptb]
\begin{center}
\includegraphics[
height=3.2768in,
width=2.9343in
]%
{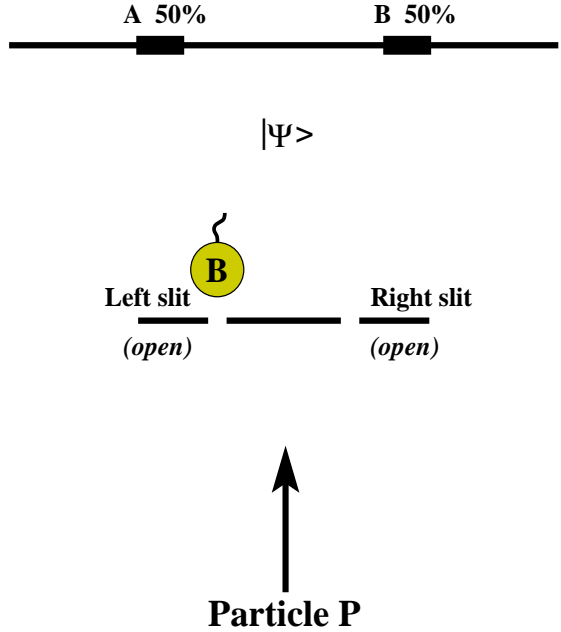}%
\caption{Variant of the Elitzur-Vaidman experiment: a bomb is placed just
after the left slit. }%
\end{center}
\end{figure}

If, instead of the bomb we put a fake bomb (referred to as the dud bomb, which
has the very same aspect of the real functioning bomb but does not interact
\emph{at all} with the particle $P$), the wave function of the system is given
by%
\begin{equation}
\left\vert \Psi\right\rangle =\frac{1}{\sqrt{2}}\left(  \left\vert
L\right\rangle +\left\vert R\right\rangle \right)  \left\vert B_{0}%
^{dud}\right\rangle =\left\vert A\right\rangle \left\vert B_{0}^{dud}%
\right\rangle \label{fake}%
\end{equation}
where $\left\vert B_{0}^{dud}\right\rangle $ describes the wave function of
the dud bomb. In this case, there is interference and the particle $P$
\emph{always} ends up in $A$.

Then, the amusing part comes: if we do not know if the bomb is a dud or not,
we can --in some but not all cases-- find out by placing it in front of the
left slit. If there is no explosion and the particle ends up in $B,$ we deduce
\emph{for sure} that the bomb is real. Namely, this outcome is not possible
for a dud, see Eq.~(\ref{fake}). Note, we have deduced that the bomb is `good'
without making it explode (that would be easy: just send the particle $P$
toward the bomb, if it goes `boom' it \emph{was} real). This situation occurs
in $25\%$ of cases in which a functioning bomb is placed behind the slit, see
Eq.~(\ref{bomb}): we can immediately `save' $25\%$ of the good bombs.
Conversely, in $50\%$ of cases the good bomb simply explodes and we lose it
(then, the particle $P$ goes to either $A$ (25\%) or to $B$ (25\%)). In the
remaining $25\%$ the good bomb does not explode, but the particle $P$ hits
$A.$ Then, we simply do not know if the bomb is good or fake: this situation
is compatible with both hypotheses. We can, however, repeat the experiment: in
the end, we will be able to save $1/4$ $+1/4\cdot1/4+...=1/3$ of the
functioning bombs.

\subsubsection{The idler particle and the delayed choice experiment}

Another interesting configuration is obtained by assuming that a second
entangled particle, denoted as $I$ (for idler), is emitted when $P$ goes
through the slit(s). The system is built in the following way: if the particle
$P$ goes through the left slit, the particle $I$ is described by the state
$\left\vert I_{L}\right\rangle $. Similarly, when the particle $P$ goes to the
right slit, the particle $I$ is described by the state $\left\vert
I_{R}\right\rangle $. We assume that the two idler states are orthogonal:
$\left\langle I_{L}|I_{R}\right\rangle =0.$ This situation resembles closely
that of delayed choice experiments \cite{dc1,dc2}.

When both slits are open the whole wave function of the system is given by:%
\begin{align}
\left\vert \Psi\right\rangle  &  =\frac{1}{\sqrt{2}}\left(  \left\vert
L\right\rangle \left\vert I_{L}\right\rangle +\left\vert R\right\rangle
\left\vert I_{R}\right\rangle \right) \nonumber\\
&  =\text{ }\frac{1}{2}\left[  \left\vert A\right\rangle (\left\vert
I_{R}\right\rangle +\left\vert I_{L}\right\rangle )+\left\vert B\right\rangle
(\left\vert I_{R}\right\rangle -\left\vert I_{L}\right\rangle )\right]  \text{
.} \label{idler}%
\end{align}
The particle $I$ is entangled with $P$, but being the latter a microscopic
object, we surely cannot apply the collapse hypothesis because the particle
$I$ is \emph{not} a measuring apparatus.

Do we have interference on the screen $S$ in this case? The answer is clear:
no. The states $\left\vert A\right\rangle \left\vert I_{L}\right\rangle ,$
$\left\vert A\right\rangle \left\vert I_{R}\right\rangle ,$ $\left\vert
B\right\rangle \left\vert I_{L}\right\rangle ,$ $\left\vert B\right\rangle
\left\vert I_{R}\right\rangle $ represent a basis of this system, thus the
probability to obtain $\left\vert A\right\rangle $ (that is, the probability
of $P$ hitting $S$ in $A$) is $1/4+1/4=1/2.$ So for $B.$ The presence of the
entangled idler state destroys the interference on $S.$

It is sometime stated that this result is a consequence of the fact that the
state of the idler particle $I$ carries the information of which way $P$ has
followed. For this reason, the interference has disappeared (this is a modern
reformulation of the complementarity principle). However, such expressions,
although appealing, are often too vague and need to be taken with care.

\emph{ }As a next step we study what happens if we perform a measurement on
the particle $I$. We study separately two distinct types of measurements.

\paragraph{Measuring $I$ in the $\left\vert I_{L}\right\rangle $-$\left\vert
I_{R}\right\rangle $ basis.}

First, we perform a measurement which tells us if the state of the idler
particle is $\left\vert I_{L}\right\rangle $ or $\left\vert I_{R}\right\rangle
.$ For simplicity, we apply the collapse hypothesis (as usual, the results
would \emph{not} change by keeping track of the whole unitary quantum
evolution). But first, we have to clarify the following issue: when do we
perform the measurement on $I$? We have two possibilities:

\begin{itemize}
\item If we measure the state of $I$ \emph{before} the particle $S$ hits the
screen, the wave function reduces to $\left\vert L\right\rangle \left\vert
I_{L}\right\rangle $ or to $\left\vert R\right\rangle \left\vert
I_{R}\right\rangle $ with $50\%$ probability, respectively. Then, the screen
$S$ performs a second measurement: we find -as usual- 50\% of times $A$
($25\%$ $\left\vert A\right\rangle \left\vert I_{L}\right\rangle $ and $25\%$
$\left\vert A\right\rangle \left\vert I_{R}\right\rangle $) and 50\% of times
$B$ ($25\%$ $\left\vert B\right\rangle \left\vert I_{L}\right\rangle $ and
$25\%$ $\left\vert B\right\rangle \left\vert I_{R}\right\rangle $).

\item If, instead, the particle $P$ arrives \emph{first} on the screen $S$,
the quantum state collapses into $\left\vert A\right\rangle (\left\vert
I_{R}\right\rangle +\left\vert I_{L}\right\rangle )/\sqrt{2}$ in $50\%$ of
cases ($A$ has clicked), or into $\left\vert B\right\rangle (\left\vert
I_{R}\right\rangle -\left\vert I_{L}\right\rangle )/\sqrt{2}$ in the other
$50\%$ of cases ($B$ has clicked). The subsequent measurement of the $I$
particle will then give $\left\vert I_{L}\right\rangle $ or $\left\vert
I_{R}\right\rangle $ ($50\%$ each).
\end{itemize}

In conclusion, we realize that it is absolutely \emph{not} relevant which
experiment is done before the other. In particular, for us sitting on the
screen $S$, it does not matter at all when and if the measurement of the idler
state is performed. We simply see no interference.

\paragraph{Measuring $I$ in the $(\left\vert I_{R}\right\rangle +\left\vert
I_{L}\right\rangle )/\sqrt{2}$-$(\left\vert I_{R}\right\rangle -\left\vert
I_{L}\right\rangle )/\sqrt{2}$ basis.}

Being the particle $P$ entangled with another particle and not with a
macroscopic state, we can also decide to perform a different kind of
measurement on $I.$ For instance, we can put a detector measuring $I$ by
projecting onto the basis $(\left\vert I_{R}\right\rangle +\left\vert
I_{L}\right\rangle )/\sqrt{2}$ and $(\left\vert I_{R}\right\rangle -\left\vert
I_{L}\right\rangle /\sqrt{2}$. If we do this measurement before the particle
$P$ has hit the screen $S,$ we have the following outcome as a consequence of
the collapse induced by the $I$-detector:%
\begin{align}
\left\vert \Psi\right\rangle  &  =\left\vert A\right\rangle (\left\vert
I_{R}\right\rangle +\left\vert I_{L}\right\rangle )/\sqrt{2}\text{ with prob.
50\% ;}\label{mes1}\\
\left\vert \Psi\right\rangle  &  =\left\vert B\right\rangle (\left\vert
I_{R}\right\rangle -\left\vert I_{L}\right\rangle )/\sqrt{2}\text{ with prob.
50\% .} \label{mes2}%
\end{align}
In the former case, the particle $P$ will surely hit $S$ in $A,$ in the latter
in $B.$

One sometimes interpret the experiment in the following way: the detector
measuring the state of $I$ as being either $(\left\vert I_{R}\right\rangle
+\left\vert I_{L}\right\rangle )/\sqrt{2}$ or $(\left\vert I_{R}\right\rangle
-\left\vert I_{L}\right\rangle /\sqrt{2}$ `erases the which-way information'.
When the detector measures $(\left\vert I_{R}\right\rangle +\left\vert
I_{L}\right\rangle )/\sqrt{2}$ we still have interference and we see the
particle $P$ in the position $A$, just as the case with two open slits
(Fig.~2). In the other case, when the detector measures $(\left\vert
I_{R}\right\rangle -\left\vert I_{L}\right\rangle )/\sqrt{2}$, we also have a
kind of interference in which the final position $B$ is the only outcome. In
the language of Ref. \cite{dc1}, one speaks of `fringes' in the former case,
and of `anti-fringes' in the latter.

However, care is needed: for us sitting on $S$, if we do not know which
measurement is performed on $I,$ we simply see that \emph{no} interference
occurs ($50\%$-$A$ and $50\%$-$B$). But, if we could then speak with a
colleague working with the $I$-detector, we would realize that, each time we
have measured $A$ he has found the state $(\left\vert I_{R}\right\rangle
+\left\vert I_{L}\right\rangle )/\sqrt{2},$ while each time we have measured
$B$ he has found $(\left\vert I_{R}\right\rangle -\left\vert I_{L}%
\right\rangle /\sqrt{2}.$ Thus, we have a \emph{correlation} of our results
(measurement of the screen $S)$ with those of the $I$-detector. This is
actually no surprise if we look at the quantum state of Eq.~(\ref{idler}).
This statement is indeed more precise than the statement of having
interference because we have erased the which-way information. Namely, we do
\emph{not} have interference.

Indeed, we can perform the measurement of $I$ even after (in principle much
time after) the screen $S$ has measured $P$ in either $A$ or $B.$ Here the
name `delayed choice' comes from: we choose if we retain the which-way
information or not. Still, the result is the same because there is no
influence on the time-ordering of the measurements. If the measurement of the
screen\ $S$ occurs first, we have a collapse onto the very same Eqs.
(\ref{mes1})-(\ref{mes2}). Then, a measurement of the idler particle $I$ would
simply find either $(\left\vert I_{R}\right\rangle +\left\vert I_{L}%
\right\rangle )/\sqrt{2}$ (correlated with $A$) and $(\left\vert
I_{R}\right\rangle -\left\vert I_{L}\right\rangle )/\sqrt{2}$ (correlated with
$B$). For sure, there is no change of the past by a measurement of the idler
state, but simply a correlation of states. Still, such a very interesting
setup visualizes many of the peculiarities of QM and can be used for quantum cryptography.

\subsubsection{Realizations of the setup}

In a two-slit experiment all the peculiarities of QM are evident due to the
fact that the particle $P$ follows (at least) two paths at the same time. This
is extremely fascinating as well as counterintuitive for our imagination based
on a childhood with rolling `classical' marbles. However, as already mentioned
in Sec. 1.1, a simple implementation of the two-slit experiment does not
produce only two possible outcomes, but gives rise to a superposition of waves
with many maxima and minima. In the following we present two possible
realizations of our Gedankenexperiment which do not make use of slits.

An interference experiment in which only two outcomes are possible can be
realized by using particles with spin $1/2$ (such as electrons in a
Stern-Gerlach-type experiment) or photons (spin 1, but due to gauge invariance
only two polarizations are realized). Clearly, all the QM features do not
depend on which particle or on which quantum number are implemented, but
solely on the presence of superpositions and on the effect of measurements. In
the case of photon polarizations we can use the fact that a photon can be
horizontally or vertically polarized (corresponding to the kets $\left\vert
h\right\rangle $ and $\left\vert v\right\rangle $ respectively). In our
analogy, the state $\left\vert h\right\rangle $ corresponds to the state of
our particle $P$ coming out from the left slit, $\left\vert h\right\rangle
\equiv\left\vert L\right\rangle ,$ and similarly $\left\vert v\right\rangle $
from the right slit, $\left\vert v\right\rangle \equiv\left\vert
R\right\rangle .$ Then, we place a detector which acts as the screen $S$ by
making a measurement in the basis $\left\vert A\right\rangle =(\left\vert
v\right\rangle +\left\vert h\right\rangle )/\sqrt{2}$ and $\left\vert
B\right\rangle =(\left\vert v\right\rangle -\left\vert h\right\rangle
)/\sqrt{2}$. In addition, we can place a second detector which plays the role
of the detector $D$ by measuring the polarization in the $\left\vert
h\right\rangle $-$\left\vert v\right\rangle $ basis. Indeed, in this case we
do not need to send the photons along two different paths, because the
polarization d.o.f. is enough for our purposes.%

\begin{figure}
[ptb]
\begin{center}
\includegraphics[
height=1.7443in,
width=4.4962in
]%
{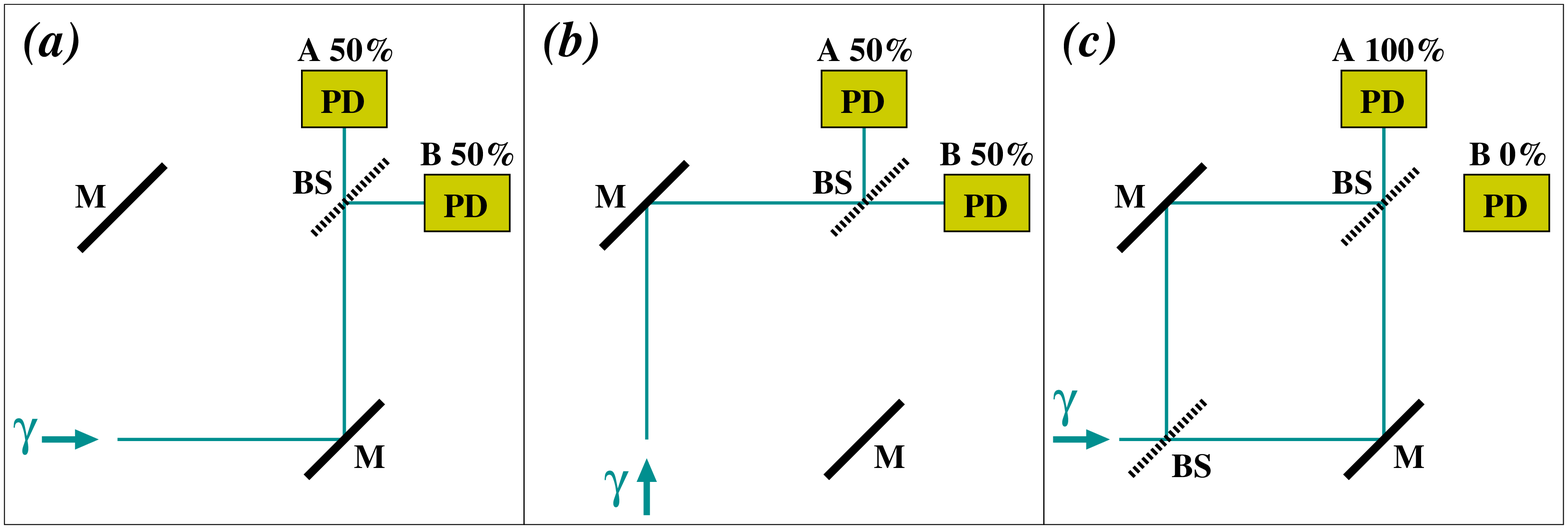}%
\caption{The Mach-Zehnder interferometer. M stands for mirror, BS for beam
splitter, and PD for photon detector. The case (a) is analogous to having only
the left slit open (Fig.~1, left side), (b) to only the right slit open
(Fig.~1, right side), (c) to both slits open (Fig.~2).}%
\end{center}
\end{figure}

Another possible realization of our setup is the Mach-Zehnder interferometer
\cite{mz}, see Fig.~5, which makes use of beam splitters. When a photon is
sent to the path of Fig.~5.a (denoted as path-1), both photon counters $A$ and
$B$ can detect the photon with a probability of 50\% . For our analogy, we
have $\left\vert path\text{-1}\right\rangle \equiv\left\vert L\right\rangle $.
Similarly, when the photon is sent to the path of Fig.~5.b (path-2), we hear a
click in $A$ or in $B$ with 50\% probability. For the analogy: $\left\vert
path\text{-2}\right\rangle \equiv\left\vert R\right\rangle $. When a beam
splitter is put in the beginning of the setup, after the photon passes
through, we get a superposition $(\left\vert path\text{-1}\right\rangle
+\left\vert path\text{-2}\right\rangle )/\sqrt{2}$ (Fig.~5.c). The inclusion
of the detector $D$, the bomb, entangled particle(s) as well as the
environment can be easily carried out.

In the end, notice that Mach-Zehnder interferometers can be constructed by
using neutrons instead of photons. The so-called neutron interferometers (see
the recent review paper \cite{neutron} and refs. therein) can be very well
controlled and allow to experimentally study quantum systems to a great level
of accuracy.

\section{Collapse vs no-collapse: there is a difference}

In this section we show that there is a difference between the collapse and
no-collapse scenarios. To this end, instead of having a detector, a bomb, or
an idler entangled state, we assume that the space between the slits and the
screen is not the vacuum. Then, we study the time evolution of the
\emph{environment }which interacts with the particle $P$. This interaction is
assumed to be soft enough not to absorb or kick away the particle in such a
way that the final outcomes on the screen $S$ are still the endpoints $A$ or
$B$.

Before the particle $P$ goes through the slit(s), the environment is described
by the state $\left\vert E_{0}\right\rangle $. First, we study the case in
which only the left slit is open. Denoting as $t=0$ the time at which $P$
passes through the left slit, the wave function of the environment evolves as
function of time $t$ as%
\begin{equation}
\left\vert \Psi(t)\right\rangle =\left\vert L\right\rangle \left\vert
E_{L}(t)\right\rangle \text{ ,}%
\end{equation}
where by construction $\left\vert E_{L}(0)\right\rangle =\left\vert
E_{0}\right\rangle $. Similarly, if only the right slit is open, at the time
$t$ the system is described by $\left\vert \Psi(t)\right\rangle =\left\vert
R\right\rangle \left\vert E_{R}(t)\right\rangle $ with $\left\vert
E_{R}(0)\right\rangle =\left\vert E_{0}\right\rangle $.

We now turn to the case in which both slits are open. It is important to
stress that, by assuming a weak interaction of the particle $P$ with the
environment, we surely do not have -at first- a collapse of the wave function,
but an evolution of the whole quantum state given by:%
\begin{align}
\left\vert \Psi(t)\right\rangle  &  =\frac{1}{\sqrt{2}}\left(  \left\vert
L\right\rangle \left\vert E_{L}(t)\right\rangle +\left\vert R\right\rangle
\left\vert E_{R}(t)\right\rangle \right) \nonumber\\
&  =\frac{1}{2}\left[  \left\vert A\right\rangle (\left\vert E_{R}%
(t)\right\rangle +\left\vert E_{L}(t)\right\rangle )+\left\vert B\right\rangle
(\left\vert E_{R}(t)\right\rangle -\left\vert E_{L}(t)\right\rangle )\right]
\text{ .} \label{psioft}%
\end{align}
This is indeed very similar to the detector case, but there is a crucial
aspect that we now take into consideration. The state $\left\vert
E_{L}(t)\right\rangle $ and $\left\vert E_{R}(t)\right\rangle $ coincide at
$t=0$ and then \emph{smoothly} depart from each other. At the time $t$ we
assume to have
\begin{equation}
c(t)=\left\langle E_{L}(t)|E_{R}(t)\right\rangle =e^{-\lambda t}\text{ .}
\label{coft}%
\end{equation}
(where $c(t)$ is taken to be real for simplicity). This is nothing else than a
gradual decoherence process. The states of the environment entangled with
$\left\vert L\right\rangle $ and $\left\vert R\right\rangle $ overlap less and
less by the time passing. The constant $\lambda$ describes the speed of the
decoherence and depends on the number of particles involved and the intensity
of the interaction. Note, strictly speaking, this non-orthogonality is also
present in the case of the detector (if no collapse is assumed), but the
overlap is amazingly small, see the estimate in Eq.~(\ref{overlap}). (In the
case of the detector $D$ of Sec. 2.2, $\lambda$ is very large and consequently
$\lambda^{-1}$ is a very short time scale, shorter than any other time scale
in the setup of Fig.~3. For that reason we assumed that the detector state
evolved for all practical purposes instantaneously from the ready-state
(pointer up) to pointing either to the left or to the right.)

Now we ask the following question: what is the probability that the particle
$P$ hits the screen in $A$? We assume that the particle $P$ hits the screen at
the time $\tau.$ At this instant, the state is given by $\left\vert \Psi
(\tau)\right\rangle $ with $\left\langle E_{L}(\tau)|E_{R}(\tau)\right\rangle
=c(\tau)$.

We now present the mathematical steps leading to $p[A,\tau]$, which, although
still simple, are a bit more difficult than the previous ones. The reader who
is only interested in the result can go directly to Eq.~(\ref{pad}).

At the time $\tau$ we express the state $\left\vert E_{L}(\tau)\right\rangle $
as%
\begin{equation}
\left\vert E_{L}(\tau)\right\rangle =c(\tau)\left\vert E_{R}(\tau
)\right\rangle +\sum_{\alpha}b_{\alpha}(\tau)\left\vert E_{R,\perp}^{\alpha
}(\tau)\right\rangle
\end{equation}
where the summation over $\alpha$ includes all states of the environment which
are orthogonal to $\left\vert E_{R}(\tau)\right\rangle $: $\left\langle
E_{R,\perp}^{\alpha}(\tau)|E_{R}(\tau)\right\rangle =0$. This expression is
possible because the set $\{E_{R}(\tau),E_{R,\perp}^{\alpha}(\tau)\}$
represents a orthonormal basis for the environment state. Its explicit
expression is be extremely complicated, but we do not need to specify it. The
normalization of the state $\left\vert E_{L}(\tau)\right\rangle $ implies that%
\begin{equation}
\left\vert c(\tau)\right\vert ^{2}+\sum_{\alpha}\left\vert b_{\alpha}%
(\tau)\right\vert ^{2}=1. \label{norm}%
\end{equation}
Then, the state of the system at the instant $\tau$ is given by the
superposition
\begin{align}
\left\vert \Psi(\tau)\right\rangle  &  =\frac{1}{2}\left[  1+c(\tau)\right]
\left\vert A\right\rangle \left\vert E_{R}(\tau)\right\rangle +\frac{1}%
{2}\left\vert A\right\rangle \sum_{\alpha}b_{\alpha}(\tau)\left\vert
E_{R,\perp}^{\alpha}(\tau)\right\rangle \nonumber\\
&  +\frac{1}{2}\left[  1-c(\tau)\right]  \left\vert B\right\rangle \left\vert
E_{R}(\tau)\right\rangle +\frac{1}{2}\left\vert B\right\rangle \sum_{\alpha
}b_{\alpha}(\tau)\left\vert E_{R,\perp}^{\alpha}(\tau)\right\rangle \text{ .}%
\end{align}
At the time $\tau$ the probability of the particle $P$ hitting $A$ is given by%
\begin{equation}
p[A,\tau]=\frac{1}{4}\left\vert 1+c(\tau)\right\vert ^{2}+\frac{1}{4}%
\sum_{\alpha}\left\vert b_{\alpha}(\tau)\right\vert ^{2}=\frac{1}{4}\left\vert
1+c(\tau)\right\vert ^{2}+\frac{1}{4}\left(  1-\left\vert c(\tau)\right\vert
^{2}\right)  \text{ ,}%
\end{equation}
where in the last step we have used Eq.~(\ref{norm}). A simple calculation
leads to%
\begin{equation}
p[A,\tau]=\frac{1}{2}+\frac{1}{2}c(\tau)=\frac{1}{2}+\frac{1}{2}%
e^{-\lambda\tau}\text{ }. \label{pad}%
\end{equation}
A similar calculation leads to the probability of the particle $P$ hitting $S$
in $B$ as%
\begin{equation}
p[B,\tau]=\frac{1}{2}-\frac{1}{2}c(\tau)=\frac{1}{2}-\frac{1}{2}%
e^{-\lambda\tau}\text{ .}%
\end{equation}
We see that `a bit' of interference is left (no matter how large the time
interval $\tau$ is):
\begin{equation}
p[A,\tau]-p[B,\tau]=e^{-\lambda\tau}\text{ ,}%
\end{equation}
showing that there is always an (eventually very slightly) enhanced
probability to see the particle in $A$ rather than in $B.$

Notice that the very same result is found by using the reduced statistical
operator:
\begin{align}
\hat{\rho}_{red}(\tau)  &  =\left\langle E_{R}(\tau)\left\vert \hat{\rho}%
(\tau)\right\vert E_{R}(\tau)\right\rangle +\sum_{\alpha}\left\langle
E_{R,\perp}^{\alpha}(\tau)\left\vert \hat{\rho}(\tau)\right\vert E_{R,\perp
}^{\alpha}(\tau)\right\rangle \nonumber\\
&  =\left(
\begin{array}
[c]{cc}%
\left\vert A\right\rangle  & \left\vert B\right\rangle
\end{array}
\right)  \left(
\begin{array}
[c]{cc}%
p[A,\tau] & c(\tau)\\
c(\tau) & p[B,\tau]
\end{array}
\right)  \left(
\begin{array}
[c]{c}%
\left\langle A\right\vert \\
\left\langle B\right\vert
\end{array}
\right)  \text{ }%
\end{align}
where $\hat{\rho}(\tau)=\left\vert \Psi(\tau)\right\rangle \left\langle
\Psi(\tau)\right\vert $. The diagonal elements are the usual Born
probabilities, while the non-diagonal elements quantify the overlap of the two
branches and become very small for increasing time. (A related subject to the
quantum evolution described here is that of the weak measurement, in which the
`measurement' is performed by a weak interaction and thus a unitary evolution
of the whole system is taken into account, see the recent review
\cite{weakquanta} and refs. therein.)

All these considerations do not require any collapse of the wave function due
to the environment (see also Ref. \cite{pascaziocollapse}). Indeed, if we
replace the environment with the detector $D$ of Sec. 2 (which was nothing
else than a particular environment), the whole discussion is still valid (but
see the comments on time scale after Eq.~(\ref{coft})). The only point when
the Born rule enters is when we see the particle being either in $A$ or in
$B,$ but -as we commented previously- in this no-collapse MWI scenario, we
\emph{do not know why} the Born rule applies \cite{rae,hsu}. In this sense,
decoherence alone is not a solution of the measurement problem \cite{adler}.
The wave function is still a superposition of different and distinguishable
macroscopic states. Still, because of decoherence, these states (branches)
become almost orthogonal, thus decoherence is an important element of the MWI
although it does not explain the emergence of probabilities.

What do theories with the collapse of the wave function predict? As long as
few particles of the environment are involved (i.e., at small times), for sure
we do not have any collapse and the entanglement in Eq.~(\ref{psioft}) is the
correct description of the system. Namely, we know that interference effects
occur for systems which contains about $1000$ (and even more) particles
\cite{orgmol}. But, if we wait long enough we can reach a critical number of
particles at which the collapse takes place. Thus, simplifying the discussion
as much as possible, according to collapse models there should be a critical
time-interval $\tau^{\ast}$ at which the probability $p[A,\tau]$
\emph{suddenly} jumps to $1/2$ \cite{footnote}:%
\begin{equation}
p[A,\tau]=\left\{
\begin{array}
[c]{c}%
\frac{1}{2}+\frac{1}{2}e^{-\lambda\tau}\text{ for }\tau<\tau^{\ast}\text{ ;}\\
\\
\frac{1}{2}\text{ for }\tau\geq\tau^{\ast}\text{ .}%
\end{array}
\right.  \label{coll}%
\end{equation}
Indeed, such a sudden jump is an oversimplification, but is enough for our
purposes: it shows that a new phenomenon, the collapse, takes place. In Fig.~6
we show schematically the difference between the `no-collapse' and the
`collapse' cases. Obviously, if $\tau^{\ast}$ is very large, it becomes
experimentally very difficult to distinguish the two curves, but the
qualitative difference between them is clear.%

\begin{figure}
[ptb]
\begin{center}
\includegraphics[
height=2.5002in,
width=3.7014in
]%
{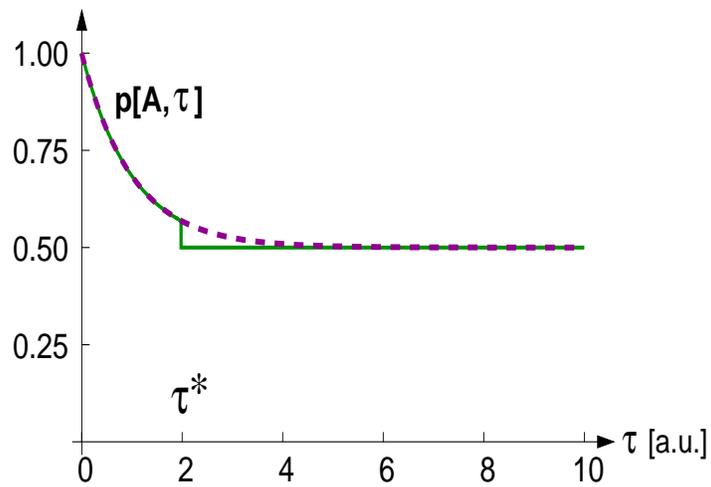}%
\caption{The quantity $p[A,\tau]$ is plotted as function of $\tau.$ The dashed
line represents the prediction of the unitary evolution of Eq.~(\ref{psioft}).
The solid line represents the prediction of the collapse hypothesis of
Eq.~(\ref{coll}): if the detection of the screen takes place for $\tau$ larger
than the critical value $\tau^{\ast}$, the state has collapsed to either $A$
or $B,$ therefore $p[A,\tau>\tau^{\ast}]=1/2$. Note, we use arbitrary
units.\ The choice of $\tau^{\ast}$ is also arbitrary and serves to visualize
the effect (it is expected to be much larger in reality).}%
\end{center}
\end{figure}

In\ Ref. \cite{hasselbach} the gradual appearance of decoherence due to
interaction of electrons with image charges has been experimentally observed.
This is analogous to our Eq.~(\ref{pad}). (For other decoherence experiment
see Ref. \cite{schlosshauershort} and refs. therein.) Indeed, it would be very
interesting to study decoherence in a setup with only two outcomes, for
instance with the help of a Mach-Zehnder interferometer or by using neutron
interferometers. Namely, even if the distinction between collapse/non-collapse
is not yet reachable \cite{bassi}, a clear demonstration of decoherence and
the experimental verification of Eq.~(\ref{pad}) would be useful on its own.

As a last step, we show that the behavior $p[A,t]=1/2$ $\forall t>\tau^{\ast}$
is a peculiarity of the collapse approach which is \emph{impossible} if only a
unitary evolution is taken into account. The proof makes use of the
Hamiltonian $H$ of the whole system (particle+slits+environment), for which we
assume that $\left\langle R\left\vert H\right\vert L\right\rangle
=\left\langle L\left\vert H\right\vert R\right\rangle =0$, i.e. the full
Hamiltonian does \emph{not} mix the states $\left\vert L\right\rangle $ and
$\left\vert R\right\rangle $. (This is indeed a quite general assumption for
the type of problems that we study: once the particle has gone through the
left slit, its wave function is $\left\vert L\right\rangle $ and stays such
(and viceversa for $\left\vert R\right\rangle ).$ Similarly, in the example of
a (photon or neutron) Mach-Zehnder interferometer, after the first
beam-splitter the path is either the lower or the upper and the whole
Hamiltonian does not mix them.) It then follows that:%
\begin{align}
\left\vert \Psi(t)\right\rangle  &  =e^{-iHt}\frac{1}{\sqrt{2}}\left(
\left\vert L\right\rangle \left\vert E_{0}\right\rangle +\left\vert
R\right\rangle \left\vert E_{0}\right\rangle \right) \nonumber\\
&  =\frac{1}{\sqrt{2}}\left(  \left\vert L\right\rangle e^{-iH_{L}t}\left\vert
E_{0}\right\rangle +\left\vert R\right\rangle e^{-iH_{R}t}\left\vert
E_{0}\right\rangle \right)
\end{align}
where we have expressed $\left\vert E_{L}(t)\right\rangle =e^{-iH_{L}%
t}\left\vert E_{0}\right\rangle $ and $\left\vert E_{R}(t)\right\rangle
=e^{-iH_{R}t}\left\vert E_{0}\right\rangle $ by introducing the Hamiltonians
$H_{L}=\left\langle L\left\vert H\right\vert L\right\rangle $ and
$H_{R}=\left\langle R\left\vert H\right\vert R\right\rangle $ which act in the
subspace of the environment. (These expressions hold because $H^{n}\left\vert
L\right\rangle \left\vert E_{0}\right\rangle =\left\vert L\right\rangle
H_{L}^{n}\left\vert E_{0}\right\rangle $ for each $n$). The overlap $c(t)$
defined in\ Eq.~(\ref{coft}) can be formally expressed as
\begin{equation}
c(t)=\left\langle E_{L}(t)|E_{R}(t)\right\rangle =\left\langle E_{0}\left\vert
e^{-i(H_{R}-H_{L})t}\right\vert E_{0}\right\rangle \text{ .}%
\end{equation}
Being $H_{L}$ and $H_{R}$ Hermitian, also $H_{R}-H_{L}$ is such. For a finite
number of degrees of freedom of the system, the quantity $c(t)$ shows a
(almost) periodic behavior and returns (very close) to the initial value $1$
in the so-called Poincar\'{e} duration time (which can be very large for large
systems). It is then excluded that $c(t)$ vanishes for $t>\tau^{\ast}.$ (At
most, it can vanish for certain discrete times, see Sec. 4, but not
continuously). Even in the limit of an infinite number of states, the quantity
$c(t)$ does not vanish but approaches smoothly zero for $t\rightarrow\infty$.

\section{Entanglement with a non-orthogonal idler state}

As a last example, we design an ideal setup in which the environment is
represented again by a single particle, the idler state (see Sec. 2.3.2).
However, we assume now that a time-evolution of the idler state takes place:
\begin{equation}
\left\vert \Psi(t)\right\rangle =\frac{1}{\sqrt{2}}\left(  \left\vert
L\right\rangle \left\vert E_{L}(t)\right\rangle +\left\vert R\right\rangle
\left\vert E_{R}(t)\right\rangle \right)  \text{ ,}%
\end{equation}
with the `environment' states now expressed in terms of the orthonormal
idler-basis $\{\left\vert I_{1}\right\rangle ,\left\vert I_{2}\right\rangle
\}$.
\begin{align}
\left\vert E_{L}(t)\right\rangle  &  =\left\vert I_{1}\right\rangle \text{
,}\\
\left\vert E_{R}(t)\right\rangle  &  =\cos(\omega t)\left\vert I_{1}%
\right\rangle +\sin(\omega t)\left\vert I_{2}\right\rangle \text{ .}%
\end{align}
Thus, while $\left\vert E_{L}(t)\right\rangle =\left\vert I_{1}\right\rangle $
is a constant over time, we assume that $\left\vert E_{R}(t)\right\rangle $
rotates in the space spanned by $\left\vert I_{1}\right\rangle $ and
$\left\vert I_{2}\right\rangle .$ Then, we can rewrite $\left\vert
\Psi(t)\right\rangle $ as
\begin{align}
\left\vert \Psi(t)\right\rangle  &  =\frac{1}{2}\left\vert A\right\rangle
\left[  \left(  1+\cos(\omega t)\right)  \left\vert I_{1}\right\rangle
+\sin(\omega t)\left\vert I_{2}\right\rangle \right] \nonumber\\
&  +\frac{1}{2}\left\vert B\right\rangle \left[  \left(  -1+\cos(\omega
t)\right)  \left\vert I_{1}\right\rangle +\sin(\omega t)\left\vert
I_{2}\right\rangle \right]  \text{ .} \label{rotidler}%
\end{align}
The probability $p[A,\tau]$ is given by%
\begin{equation}
p[A,\tau]=\frac{1}{2}+\frac{1}{2}\cos(\omega\tau)
\end{equation}
where $\tau$ is the time at which the particle $P$ hits the screen.%

\begin{figure}
[ptb]
\begin{center}
\includegraphics[
height=2.4613in,
width=4.4503in
]%
{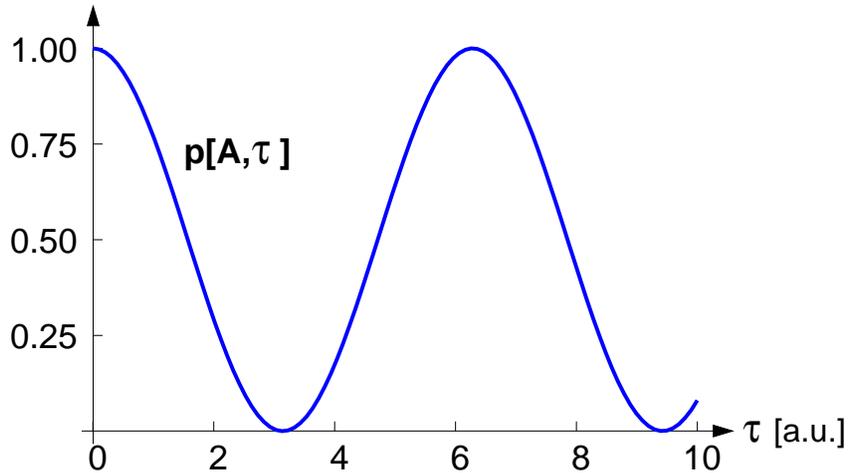}%
\caption{Quantity $p[A,\tau]$ as function of $\tau$ in the case of
entanglement with an idler state according to Eq.~(\ref{rotidler}).}%
\end{center}
\end{figure}

In conclusion, in a real implementation of this simple idea, it would be
interesting to see the appearance and the disappearance of interference (with
both fringes and antifringes) as function of the time of flight $\tau$, see
Fig.~7. It should be however stressed that the full interaction Hamiltonian
does not act on the idler state alone. Indeed, the corresponding Hamiltonian
has the form%
\begin{equation}
H=\alpha(\left\vert R\right\rangle \left\vert I_{1}\right\rangle \left\langle
R\right\vert \left\langle I_{2}\right\vert +\text{ h.c.}).
\end{equation}
This is indeed a quite peculiar type of interaction because the idler state
rotates only if the particle $P$ is in the state $\left\vert R\right\rangle $
(in the language of Sec. 4, it means: $H_{L}=0,$ $H_{R}=\alpha(\left\vert
I_{1}\right\rangle \left\langle I_{2}\right\vert +$ h.c.$).$). This implies
that the spatial trajectory of both states $\left\vert I_{1}\right\rangle $
and $\left\vert I_{2}\right\rangle $ must be the same, otherwise the overlap
$\left\langle E_{L}(t)|E_{R}(t)\right\rangle $ would be an extremely small
number and the effect that we have described would not take place.

\section{Conclusions}

We have presented an ideal interference experiment in which we have compared
the unitary evolution and the existence of a collapse of the wave function. We
have analyzed the case in which a detector measures the which-way information
and we have shown that the collapse postulate as well as the no-collapse
unitary evolution lead to the same outcome: the disappearance of interference
on the screen. In the unitary (no-collapse) evolution, this is true
\emph{only} if the states of the detector are orthogonal. This is surely a
very good, but not exact, approximation. It was then possible to describe
within the very same Gedankenexperiment two astonishing quantum phenomena: the
Elitzur-Vaidman bomb and the delayed-choice experiment.

We have then turned to a description of the entanglement with the environment.
The phenomenon of decoherence ensures that the interference smoothly
disappears. However, as long as the quantum evolution is unitary, it never
disappears completely. Conversely, the \emph{real} collapse of the wave
function introduces a new kind of dynamics which is not part of the linear
Schr\"{o}dinger equation. While the details differ according to which model is
chosen \cite{bassi}, the main features are similar: a quantum state in which
one has a delocalized object (superposition of `here' and `there') is
\emph{not} a stable configuration, but is metastable and decays to a definite
position (either `here' or `there'). In conclusion, the collapse and the
no-collapse views are intrinsically different, as Fig.~5 shows. At a
fundamental level, the unitary (no-collapse) evolution leads quite naturally
to the many worlds interpretation in which also detectors and observers are
included in a superposition (for a different view see the Bohm interpretation
\cite{bohm}).

Even if the distinction between the collapse and the no-collapse alternatives
is probably still too difficult to be detected at the moment, the
demonstration of decoherence in an experiment with two final states would be
an interesting outcome on its own (see the dashed curve in Fig.~6) . Also a
situation in which an entangled particle is emitted in such a way that an
`oscillating interference' takes place (see Fig.~7) might be an interesting possibility.

A further promising line of research to test the existence of the collapse of
the wave function is the theoretical and experimental study of unstable
quantum systems. The non-exponential behavior of the survival probability for
short times renders the so-called Zeno and Anti-Zeno effects possible
\cite{misra,ghirardi,zeno,duecan}: these are modifications of the survival
probability due to the effect of the measurement, which have been
experimentally observed \cite{raizen}. The measurement of an unstable system
(for instance, the detection of the decay products) can be modelled as a
series of ideal measurements in which the collapse of the wave function
occurs, but can also be modelled through a unitary evolution in which the wave
function of the detector is taken into account and no collapse takes place
\cite{schulman,pascapulsed,koshinoprl}. Then, if differences between these
types of measurement appear, one can test how a detector is performing a
certain measurement \cite{gpinpreparation}. Quite remarkably, such effects are
not restricted to nonrelativistic QM, but hold practically unchanged also in
the context of relativistic quantum field theory \cite{zenoqft} and are
therefore applicable in the realm of elementary particles.

In conclusions, Quantum Mechanics still awaits for better understanding in the
future. It is surely of primary importance to test the validity of (unitary)
standard QM for larger and heavier bodies. In this way the new collapse
dynamics, if existent, may be discovered.

\bigskip

\textbf{Acknowledgment:} These reflections arise from a series of seminars on
`Interpretation and New developments of QM' and lectures `Decays in\ QM\ and
QFT', which took place in Frankfurt over the last 4 years. The author thanks
Francesca Sauli, Stefano Lottini, Giuseppe Pagliara, and Giorgio Torrieri for
useful discussions. Stefano Lottini is also aknwoledged for a careful reading
of the manuscript and for help in the preparation of the figures.


\begin{thebibliography}{99}                                                                                               %


\bibitem {sakurai}J. J. Sakurai, \textit{Modern Quantum Mechanics},
Addison-Wesley Publishing Company (1994).

\bibitem {omnes}R.~Omnes, \textit{Interpretation Of Quantum Mechanics,}
Princenton, Princeton University Press, 1994.


\bibitem {100}M.~Tegmark and J.~A.~Wheeler,
Wiss.\ Dossier\ \textbf{2003N1} (2003) 6] [quant-ph/0101077].


\bibitem {bell}J.~S.~Bell,
(1981) 41;
J.~S.~Bell,
(1964) 195;
J.~S.~Bell,
(1966) 447.


\bibitem {bassighirardi}
A.~Bassi and G.~C.~Ghirardi,
Phys.\ Rept.\ \textbf{379} (2003) 257 [quant-ph/0302164].


\bibitem {penrosebook}R. Penrose, \textit{The road to reality: A complete
guide to the laws of the Universe}, Vintage Book, London.




\bibitem {bassi}
A.~Bassi, K.~Lochan, S.~Satin, T.~P.~Singh and H.~Ulbricht,
Rev.\ Mod.\ Phys.\ \textbf{85} (2013) 471 [arXiv:1204.4325 [quant-ph]].


\bibitem {grw}
G.~C.~Ghirardi, O.~Nicrosini, A.~Rimini and T.~Weber,
Nuovo Cim.\ B \textbf{102} (1988) 383.


\bibitem {pearle}
P.~M.~Pearle,
Phys.\ Rev.\ D \textbf{13} (1976) 857.


\bibitem {penrose}R.~Penrose,
(1996) 581.


\bibitem {diosi}
L.~Diosi,
J.\ Phys.\ A \textbf{21} (1988) 2885.


\bibitem {singh}T.~P.~Singh,
J.\ Phys.\ Conf.\ Ser.\ \textbf{174} (2009) 012024 [arXiv:0711.3773 [gr-qc]].


\bibitem {bomb}A.~C.~Elitzur and L.~Vaidman,
\textbf{23}, Issue 7 987-997 (1993) [hep-th/9305002].


\bibitem {dc1}
Y.~-H.~Kim, R.~Yu, S.~P.~Kulik, Y.~H.~Shih and M.~.O.~Scully,
Phys.\ Rev.\ Lett.\ \textbf{84} (2000) 1 [quant-ph/9903047].


\bibitem {dc2}
S.~P.~Walborn, M.~O.~Terra Cunha, S.~Padua and C.~H.~Monken,
Phys.\ Rev.\ A \textbf{65} (2002) 033818.


\bibitem {everett}
H.~Everett,
Rev.\ Mod.\ Phys.\ \textbf{29} (1957) 454.


\bibitem {wheeler}J. A. Wheeler,
Rev. Mod. Phys. \textbf{29} (1957) 463


\bibitem {dewitt}DeWitt Bryce S,
Physics Today, Vol. 23, No. 9 (September 1970)

\bibitem {tegmark}
M.~Tegmark,
Nature \textbf{448} (2007) 23 [arXiv:0707.2593 [quant-ph]].


\bibitem {notemwi}Originally, Everett \cite{everett} introduced the concept of
`relative state formulation', which was reinterpreted as the MWI by Wheeler
and Dewitt \cite{wheeler,dewitt}. The MWI is the most natural interpretation
when no collapse is present, but the definition of what is a `world' is not
trivial. Intuitively, it is a piece of the wave function which is a
pointer-state, i.e. it does not contain spacial superpositions of macroscopic
objects. Other points of view, such as `many histories' and `many minds' were
also considered.

\bibitem {dsw}
D.~Deutsch,
quant-ph/9906015;
S. Saunders,
Proc.Royal Society A460 67-68 (2004); D. Wallace,
Studies in the History and Philosophy of Modern Physics \textbf{34}, 415-442 (2003).

\bibitem {zurek}
W.~H.~Zurek,
Phys.\ Rev.\ A \textbf{87} (2013) 5, 052111 [arXiv:1212.3245 [quant-ph]].

\bibitem {rae}A. M. Rae,
Studies in Hist Phil Modern Phys \textbf{40}, 243-250 (2009) [arXiv:0810.2657 [quant-ph]].

\bibitem {hsu}
S.~D.~H.~Hsu,
Mod.\ Phys.\ Lett.\ A \textbf{27} (2012) 1230014 [arXiv:1110.0549
[quant-ph]].


\bibitem {commentmwi}Notice that in the case of Eq.~(\ref{send}) one could
understand the MWI by noticing that there are two worlds, ergo the subjective
probability to be in one of those is $50\%$ in agreement with the Born rule.
However, this is a particular case with equal coefficients. When the
coefficients in front of the kets are \emph{not} $1/\sqrt{2}$ (but -say- $a$
and $b$ with $\left\vert a\right\vert ^{2}+\left\vert b\right\vert ^{2}=1$)
one still has two worlds but the subjective probability to be in one of those
is not $1/2,$ but the one given by the Born rule ($\left\vert a\right\vert
^{2}$ and $\left\vert b\right\vert ^{2}$ respectively). This is exactly the
point discussed in Refs. \cite{dsw,rae,hsu} with, however, different conclusions.

\bibitem {introdec}F. Marquardt and A. P\"{u}ttmann,
Lect. Notes given at Langeoog, October 2007, arXiv:0809.4403v1 [quant-ph];
Klaus Hornberger,
221-276 (2009), arXiv:quant-ph/0612118v3.


\bibitem {zurekrev}
W.~H.~Zurek,
Rev.\ Mod.\ Phys.\ \textbf{75} (2003) 715.


\bibitem {schlosshauer}
M.~Schlosshauer,
Rev.\ Mod.\ Phys.\ \textbf{76} (2004) 1267 [quant-ph/0312059].


\bibitem {schlosshauershort}M. Schlosshauer,
\textit{Compendium of Quantum Physcis: Concepts, History and Philosophy},
edited by D. Greenberger, K. Hesntschel, and F. Weinert (Springer,
Berlin/Heidelberg 2009).

\bibitem {zeilinger}G. Kwiat \textit{et al,} Phys. Rev. Lett. \textbf{74}
(1995) 4763.

\bibitem {mz}L. Zehnder,
Zeitschrift f\"{u}r Instrumentenkunde. Nr. 11, 1891, S. 275--285; L. Mach
Zeitschrift f\"{u}r Instrumentenkunde. Nr. 12, 1892, S. 89--93.

\bibitem {neutron}J. Klepp, S. Sponar, and Y. Hasegawa, to appear in Prog.
Theor. Exp. Phys. 2012, arXiv: 1407.2526 [quant-ph].

\bibitem {weakquanta}B. E. Y. Swensson, Quanta Vol \textbf{2} No 1 (2013).

\bibitem {pascaziocollapse}
M.~Namiki and S.~Pascazio,
Phys.\ Rev.\ A \textbf{44} (1991) 39.


\bibitem {adler}
S.~L.~Adler,
Stud.\ Hist.\ Philos.\ Mod.\ Phys.\ \textbf{34} (2003) 135
[quant-ph/0112095].


\bibitem {orgmol}S. Gernich \textit{et al},
Nature Communications, Vol. \textbf{2}, id. 263 (2011).

\bibitem {footnote}In the presented example we vary the time of flight $\tau$
by keeping all the rest unchanged, but the crucial point is the number of
particles involved. Alternatively, one could change the density of the
particles of the environment, which induces a change of the parameter
$\lambda.$ In that case, one would have a critical $\lambda_{\ast}$.

\bibitem {hasselbach}P. Sonnentag and F. Hasselbach, Phys. Rec. Lett. 98
200402 (2007).

\bibitem {bohm}An alternative point of view is the Bohm interpretation
(D.~Bohm,
Phys. Rev. A \textbf{85} (1952) 166) in which an equation describing the
trajectories of the particles is added (the positions are the hidden variables
of this approach). The Born rule is put in from the very beginning. An
extension of the Bohm interpretation to the relativistic framework and to
quantum field theories is a difficult task, see O. Passon,
arXiv:quant-ph/0412119
for a critical analysis.

\bibitem {misra}B.~Misra and E.~C.~G.~Sudarshan,
J.\ Math.\ Phys.\ \textbf{18} (1977) 756;
A.~Degasperis, L.~Fonda and G.~C.~Ghirardi,
Nuovo Cim.\ A \textbf{21} (1973) 471.


\bibitem {ghirardi}L.~Fonda, G.~C.~Ghirardi and A.~Rimini,
Rept.\ Prog.\ Phys.\ \textbf{41} (1978) 587; L. A. Khalfin, 1957 Zh. Eksp.
Teor. Fiz. \textbf{33} 1371. (Engl. trans. Sov. Phys. JETP \textbf{6} 1053).


\bibitem {zeno}K.~Koshino and A.~Shimizu,
Phys.\ Rept.\ \textbf{412} (2005) 191; A. G. Kofman and G. Kurizki, Nature
(London) 405, 546 (2000); P.~Facchi, H.~Nakazato, S.~Pascazio
Phys.\ Rev.\ Lett.\ \textbf{86} (2001) 2699-2703.

\bibitem {duecan}F.~Giacosa,
Found.\ Phys.\ \textbf{42} (2012) 1262 [arXiv:1110.5923 [nucl-th]];
F.~Giacosa,
Phys.\ Rev.\ A \textbf{88}, 052131 (2013) [arXiv:1305.4467 [quant-ph]];
F.~Giacosa and G.~Pagliara,
Quant.\ Matt.\ \textbf{2} (2013) 54 [arXiv:1110.1669 [nucl-th]].

\bibitem {raizen}S.~R.~Wilkinson, C.~F.~Bharucha, M.~C.~Fischer,
K.~W.~Madison, P.~R. Morrow, Q.~Niu, B.~Sundaram, M.~G.~Raizen, Nature
\textbf{387}, 575 (1997).
M.~C.~Fischer, B.~Guti{\'{e}}rrez-Medina and M.~G.~Raizen, Phys. Rev. Lett.
\textbf{87}, 040402 (2001).


\bibitem {schulman}L.S. Schulman, Phys. Rev. A \textbf{57} 1509n (1998).

\bibitem {pascapulsed}P. Facchi and S. Pascazio, Fortschritte der Physik
\textbf{49}, 941 (2001).

\bibitem {koshinoprl}K.~Koshino and A.~Shimizu,
Phys.\ Rev.\ Lett. \textbf{92} (2004) 030401.


\bibitem {gpinpreparation}
F.~Giacosa and G.~Pagliara,
arXiv:1405.6882 [quant-ph].

\bibitem {zenoqft}F.~Giacosa, G.~Pagliara,
Mod.\ Phys.\ Lett.\ \textbf{A26 } (2011) 2247-2259 [arXiv:1005.4817
[hep-ph]];
F.~Giacosa and G.~Pagliara,
Phys.\ Rev.\ D \textbf{88} (2013) 025010 [arXiv:1210.4192 [hep-ph]];
F.~Giacosa and G.~Pagliara,
Phys.\ Rev.\ C \textbf{76} (2007) 065204 [arXiv:0707.3594 [hep-ph]];
F.~Giacosa and T.~Wolkanowski,
Mod.\ Phys.\ Lett.\ A \textbf{27} (2012) 1250229 [arXiv:1209.2332 [hep-ph]].

\end{thebibliography}
\end{document}